# Anisotropic suppression of the phononic thermal conductivity by magnetic field in SmAlSi

Mujeeb Ahmad [1], Md Shahin Alam [1], Xiaohan Yao [2], Fazel Tafti [2], Marcin Matusiak [2,3] *

1. *International Research Centre MagTop, Institute of Physics, Polish Academy of Sciences, Aleja Lotnikow 32/46, PL-02668 Warsaw, Poland*
2. *Department of Physics, Boston College, Chestnut Hill, Massachusetts, 02467, USA*
3. *Institute of Low Temperature and Structure Research, Polish Academy of Sciences, ulica Okólna 2, 50-422 Wrocław, Poland*

We report the thermal and electrical conductivity data for the magnetic Weyl semimetal SmAlSi measured in a magnetic field ($B$) with two different orientations. In one case, $B$ was applied perpendicular to the heat or charge current, in the other they were parallel. For both configurations, the magnetic field affects the magnetic structure identically as $B$ is always parallel to the equivalent tetragonal axis. Our results indicate that phonon heat transport in response to the magnetic field exhibits strong anisotropy at low temperature: it appears to be independent of $B$ in the perpendicular configuration but is strongly suppressed in the parallel configuration. Understanding this unusual behaviour can lead to designing better materials for thermoelectricity or directional heat switches.

Corresponding author, email:  *m.matusiak@intibs.pl

**Introduction**

The interaction between phonons and magnetic order is not only interesting from a fundamental research point of view but could also yield potential applications. For example, the spin – lattice coupling can result in the suppression of phononic thermal conductivity [1-3], which is a desirable feature for improving thermoelectric power generation efficiency. If the attenuation of phononic thermal conduction can be controlled, for instance by a magnetic field ($B$), it becomes possible to develop thermal switching devices [4,5]. Furthermore, when the effect depends on the specific orientation of $B$ with respect to the thermal gradient, a directional switch can be designed.

Here, we investigated the thermal conductivity of SmAlSi magnetic Weyl semimetal for two orientations of the in-plane magnetic field, which was parallel or perpendicular to the thermal gradient. However, in either case $B$ was applied along the equivalent crystallographic axis due to the tetragonal structure of SmAlSi. The expected difference in response of the electronic thermal conductivity to the differently oriented magnetic field was evident, but we also discovered that the phononic contribution showed large anisotropy. Analysis of the experimental data indicates that relationship between the magnetic ordering and the phononic thermal transport exhibits remarkable anisotropy.

**Methods**

Single crystals of SmAlSi were grown by a self-flux technique, details of which were described in the previous report [6]. SmAlSi crystallizes in a non-centrosymmetric centred tetragonal structure, $I4_1md$ ($C_{4v}$) that was confirmed with powder x-ray diffraction using Bruker D8 ECO system with the FullProf suite used for the Rietveld refinement.

For the transport measurements, a rectangular bar with dimensions 2.1 x 0.79 x 0.23 mm$^3$ was cut from a suitable single crystal with the longest side of the sample oriented along the [1 0 0] direction (crystallographic $a$ – axis) and the shortest side along the [0 0 1] direction

(crystallographic $c$ – axis). The electrical resistivity ($\rho$) was measured using a four-point technique with an alternating electric current flowing along $a$. The current contacts were made on the cross-section surface rather than point-like to ensure the homogeneous current distribution and minimize extrinsic effects. The experiments were performed in the temperature ($T$) range 1.8 - 300 K and in the magnetic field ($B$) up to 14.5 T applied parallel and perpendicular to the electric or heat current, in both configurations $B \parallel a$.

The isolated heater method was used for the thermal conductivity ($\kappa$) measurements, it was described in detail in [7]. During measurements the thermal gradient ($\nabla T$) was applied along $a$-axis of the single crystal SmAlSi. A DC technique was used to measure field dependences, with up and down sweeps performed to extract the field-symmetric component of the signal. For temperature ramps at constant field the quasi-AC mode was used.

**Results**

The Weyl semimetal SmAlSi possesses an interesting magnetic phase diagram that includes Weyl-mediated spiral magnetism at low temperature [8]. On the basis of neutron diffraction, transport, and thermodynamic data four magnetic phases below $T \approx 12$ K there were identified, in addition to the so-called A phase, presumably containing topological magnetic excitations [8]. The anomalies related with transition between phases can be seen in Fig. 1, where the temperature dependences of the electrical resistivity ($\rho$) are shown for several values of the applied magnetic field ($B$). The measurements were conducted at two different magnetic field orientations, namely the perpendicular ($B_\perp \parallel a \perp j$) and parallel ($B_\parallel \parallel a \parallel j$) ones, where $a$ and $j$ denote the crystallographic $a$-axis and the electric current, respectively. It should be noted that SmAlSi has the tetragonal crystallographic structure, hence $B_\perp$ and $B_\parallel$ are directed along structurally equivalent direction even though they are perpendicular to each other. Therefore, at a given value of magnetic field both $B_\perp$ and $B_\parallel$ have the same effect on the magnetic structure. In contrast, the orientation of $B$ has substantial impact on the flow of charge carriers as depicted in inset in Fig. 1 presenting field dependences (for $B_\perp$ and $B_\parallel$) of the resistivity at $T$ = 20 K. The transverse magnetoresistance at this temperature is large and positive reaching about 350 %, whereas for the parallel configuration, with no Lorentz force involved, the resistivity initially increases slightly, but above $B_\parallel \approx 4$ T the longitudinal magnetoresistance becomes negative. This kind of behaviour can be assigned to the non-trivial topology of SmAlSi, which as a Weyl semimetal is expected to exhibit the chiral magnetic effect. This phenomenon arises from an imbalance in the number of Weyl fermions of opposite chirality in the presence of parallel electric and magnetic fields [9, 10]. The expected anomalous chiral contribution to the electrical conductivity ($\sigma_{ch}$) should depend like $B^2$ in the quasi-classical regime and for $\mu \gg T$ [11]:

$$\sigma_{ch} = N_W \frac{e^2}{8\pi^2 \hbar} \frac{(eB)^2 v^3}{\mathcal{E}_F^2} \tau_{WP}, \qquad (2)$$

where $N_W, e, v, \mathcal{E}_F, \tau_{WP}$ are: number of Weyl nodes pairs, elementary charge, Fermi velocity, Fermi energy, and inter-valley Weyl relaxation time, respectively. Figure 2 confirms that in the high field limit the conductivity (calculated as $\sigma(B) = 1/\rho(B)$) is dominated by the quadratic term, which is characteristic of the chiral anomaly. In SmAlSi this is evident over a wide range of temperatures. While there are other mechanisms that can possibly lead to negative longitudinal magnetoresistance [12,13], the measurements of the thermal conductivity ($\kappa$) can provide addition evidence that the effect is in fact stemming from non-trivial topology of the electronic structure. Such studies were performed for $Bi_{1-x}Sb_x$ [14] and NdAlSi [15] Weyl semimetals showing that anomalous chiral thermal conductivity is related to the anomalous chiral electrical conductivity in a way predicted by the Wiedemann – Franz law, i.e., $\kappa = \sigma L T$, where $L$ is the Lorenz number.

However, the electrons are not the only heat carriers in a crystal and the analysis of the thermal conductivity needs to start with quantifying other contributions. To this end, we measured $\kappa(B)$ of SmAlSi for several different temperatures with two magnetic field configurations (i.e., $B_\perp \parallel a \perp q$ and $B_\parallel \parallel a \parallel q$, where $q$ is the heat current parallel to the applied temperature gradient $\nabla T$). These are presented in Fig. 3. As could be expected from the Wiedemann – Franz law, $\kappa(B_\perp)$ decreases with field. On the other hand, behaviour of $\kappa(B_\parallel)$ is more complex. While at low temperatures $\kappa(B_\parallel)$ rises with the magnetic field in the manner inferred from the electrical conductivity field dependences shown in Fig. 2, this increase in $\kappa(B_\parallel)$ vanishes with temperature more rapidly than for the corresponding $\sigma(B_\parallel)$.

**Discussion**

When the magnetic field is oriented perpendicular to the thermal gradient the evolution of $\kappa(B_\perp)$ is simply correlated with changes of the electrical conductivity. For a given temperature $\kappa(B_\perp)$ can be calculated as $\kappa_{WF}(B_\perp) + \kappa_{ph}^\perp$, where $\kappa_{WF}(B_\perp) = \sigma(B_\perp) L T$, and $\kappa_{ph}^\perp$ is the phononic thermal conductivity. Fig. 3 compares for different temperatures such obtained magnetic field dependences of the thermal conductivity (black lines) with the respectively measured actual data (light – coloured points). An apparent very good match between the thermal and electrical-based values can be achieved with the simple assumption that $\kappa_{ph}^\perp$ is independent of $B_\perp$. This corresponds to a trivial situation, in which the magnetic field does not affect movement of chargeless phonons [14,16,17]. In SmAlSi this remains true despite the magnetic field drives the system through transitions from phase 2 to phase 4 and from phase 4 to phase 3 (see the *B-T* phase diagram of SmAlSi shown in Fig. S1 in the Supplemental Material [18]). The resulting $\kappa_{ph}^\perp(T)$ is presented in the main panel of Fig. 4, while the lower inset of Fig. 4 shows the temperature dependence of the Lorenz number. Over the temperature range studied, $L$ is almost constant at about $0.8\, L_0$, where $L_0 = \pi^2/3\, (k_B/e)^2$ is the Sommerfeld value ($k_B$ is the Boltzmann constant, $e$ is the elementary charge) expected for the Fermi liquid when charge carriers are scattered elastically. Slightly smaller than $L_0$ value of $L$ is likely related to small-angle inelastic scattering of electrons presumably present in this temperature range [19, 20].

Intriguingly, for the parallel field configuration the same approach fails to produce a satisfactory match between the measured thermal conductivity and one calculated from the electric conductivity results, i.e., $\sigma(B_{\parallel})\, L\, T + \kappa_{ph}^{\parallel}$, as long as $\kappa_{ph}^{\parallel}$ is assumed to be independent of field. However, a good correspondence between $\kappa_{WF}(B_{\parallel})$ and $\kappa(B_{\parallel})$ can be obtained once the phononic contribution is allowed to be simply linearly dependent on the magnetic field, namely $\kappa_{ph}^{\parallel} = \kappa_{ph}(0\ T) + CB$, where $C$ is a constant. A hint that we should focus on the phonon transport comes from the recent angle-resolved photoemission spectroscopy (ARPES) measurements indicated negligible coupling between localised 4f electrons and conduction electrons [21]. Using the Lorenz numbers determined from $\kappa(B_{\perp})$ we calculated field dependences of the thermal conductivity that in Fig. 3 were superimposed as yellow lines on $\kappa(B_{\parallel})$ experimental data. For the corresponding temperatures, the two dependences match each other very well, indicating that $\kappa_{ph}$ is suppressed by the magnetic field, but unusually, only when $B$ is parallel to the thermal gradient.

As SmAlSi is a Weyl semimetal, such an anisotropy in response to the magnetic field could in principle be related to the topologically nontrivial attributes of the electronic structure. Since the $B^2$ increase of both the electrical and thermal conductivities was observed in the parallel field configuration, one scenario might be that the phonon heat transport is affected by the emergence of the chirality – related anomaly. Theoretical studies indicate that the chiral magnetic effect couples with certain phonon modes [22] and that in the presence of the chiral anomaly the phonon dispersion is modified [23].

Alternatively, the attenuation of the phononic thermal conductivity in field may result from anisotropic interaction between phonons and magnetic moments of Sm ions. A hint supporting this scenario comes from the sharp drop in $\kappa_{ph}(B_{\parallel} = 14.5\ T)$ at $T \approx 7\ K$ as shown in the lower inset of Fig. 4. This is the temperature of the transition between the magnetic phases 3 and 4 (Fig. S1 [SM]), which is also confirmed by the presence of anomalies in $\rho(T)$ dependences measured at $B_\parallel = 14.5$ T and $B_\perp = 14.5$ T (Fig. 1). This decrease in $\kappa_{ph}$ is expected due to spin – phonon coupling that can lead to increased scattering and substantial reduction of the mean free path of the phonons. Such a phenomenon was studied theoretically [24, 25] and observed experimentally in magnetic materials like MnO [26] and CoF2 [27]. In HoMnO$_3$ it was demonstrated that coupling of spin fluctuations and lattice vibrations can suppress the phonon lifetimes also in the paramagnetic phase [28].

A further indication that the phonon scattering is enhanced only for the parallel configuration comes from the resemblance of the non-electronic part of the thermal conductivity ($\kappa_{ne}$) to the lattice thermal conductivity estimated from the specific heat data ($\kappa_l$). The temperature dependences of these quantities are shown in the upper inset in Fig. 4. The former is calculated as $\kappa_{ne} = \kappa - \kappa_{WF}$ with the Lorenz number estimated before ($L \approx 0.8\ L_0$), the latter is supposed to be proportional to the specific heat ($\kappa_l = 1/3\ C_v\ v_s\ l_{ph}$) under assumption that in the limited temperature range the speed of sound ($v_s$) and the mean free path do not change significantly and the constant-volume specific heat $C_v \approx C_p$. The lattice specific heat is assumed to be equal to one of non-magnetic reference material LaAlSi. Apparently, $\kappa_{ne}(T)$ for $B = 0$ T and $B_\perp = 14.5$ T change at the similar rate as $\kappa_l(T)$, which would imply that the mean free path of phonons is approximately constant, but $\kappa_{ne}(T)$ for and $B_\parallel = 14.5$ T decreases with the temperature more rapidly, suggesting the presence of additional scattering only in the parallel magnetic field configuration.

It is worth pointing out the effectiveness of the scattering, which at the transition result in the phonon mean free path close to inverse wave vector inverse of the wave vector of thermally excited phonons $q_s = \frac{k_B T}{\hbar v_s}$, where $\hbar$ is the reduced Planck constant and $v_s$ is the speed of sound. The latter is calculated to be $v_s \approx 3700$ m/s based on the crystallographic parameters of SmAlSi and the Debye temperature ($\Theta_D$) estimated from the specific heat data of isostructural LaAlSi: $v_s = \frac{k_B \Theta_D}{\hbar} \left(\frac{V}{6\pi^2 N}\right)^{1/3}$, where $V$ is the volume of the unit cell and $N$ is the number of atoms in the unit cell. At $T = 7$ K the resulting $q_s$ and $l_{ph}(B_{\parallel} = 14.5$ T) are comparable, which marks the onset of the localisation, where phonons are essentially trapped due to scattering.

Despite indications that the observed suppression of the phononic thermal conductivity is a consequence of phonon – spin interactions, it is difficult to pinpoint the specific mechanism leading to the anisotropy without knowing the exact type of magnetic ordering in the phases 1-4. If the phonon-glass-like behaviour were due to spin disorder, as reported for $EuTiO_3$ [3], it would be difficult to understand the anisotropy of the effect. The same is true for the scenario, in which the cause of the strong phonon scattering is spin fluctuation of the spin-liquid state, as suggested in case of $Tb_2Ti_2O_7$ [2]. Despite recently discussed possibility of frustrated isotropic short-range superexchange between the 4f moments in SmAlSi [29], it would again difficult to explain directionality of the phenomenon observed here.

One possible scenario may arise from recent zero-field thermal conductivity measurements in $EuIr_4In_2Ge_4$ [30]. It has been suggested that the anisotropy observed at low temperature may be due to the interaction of phonons with one-dimensional chains of short-range ordered magnetic moments of europium ions, even if $EuIr_4In_2Ge_4$ is in the paramagnetic phase. It sounds tempting to hypothesise that a similar mechanism could lead to anisotropy of the thermal conductivity in SmAlSi, if increasing magnetic field encourages formation of spin chains that interact differently with phonons moving along or across them.

**Summary**

We investigated the influence of the magnetic field on electrical and thermal transport properties of SmAlSi in two configurations. In one setup the magnetic field was applied perpendicular to the electric current or thermal gradient, while in the other configuration they were parallel. In both cases $B$ was applied along the crystallographic $a$ - axis. For the parallel configuration a $B^2$ - like increase in both electrical and thermal conductivities was observed at low temperature, consistent with the anomalous chiral current intrinsic to Weyl semimetals.

Based on differences caused by altering the magnetic field orientation, we indicated substantial change in the behaviour of the phononic thermal conductivity. Specifically, while $\kappa(B_\perp)$ shows minimal magnetic field impact on phonons, $\kappa(B_\parallel)$ exhibits strong suppression of the phononic contribution, approaching the localisation level. We attribute the observed anisotropy to the coupling between phonons and magnetic moments. A possible scenario involves induction by a magnetic field of one-dimensional spin chains whose interaction with lattice is anisotropic. Further research to reveal the specific mechanism behind this phenomenon may find application in the design of new materials for heat management.

**Acknowledgments**

We thank Piotr Stachowiak, Piotr Surówka and Francisco Pena Benitez for insightful discussions. This research was partially supported by the "MagTop" project (FENG.02.01-IP.05-0028/23) carried out within the "International Research Agendas" programme of the Foundation for Polish Science co-financed by the European Union under the European Funds for Smart Economy 2021-2027 (FENG). The work at Boston College was funded by the U.S. Department of Energy, Office of Basic Energy Sciences, Division of Physical Behavior of Materials under award number DE-SC0023124. This material is based upon work supported by the Air Force Office of Scientific Research under award number FA9550-23-1-0124.

**Competing financial interests:**

The authors declare no competing financial interests.

**M. Matusiak ORCID iD**: 0000-0003-4480-9373

**F. Tafti ORCID iD**: 0000-0002-5723-4604

**Figures**

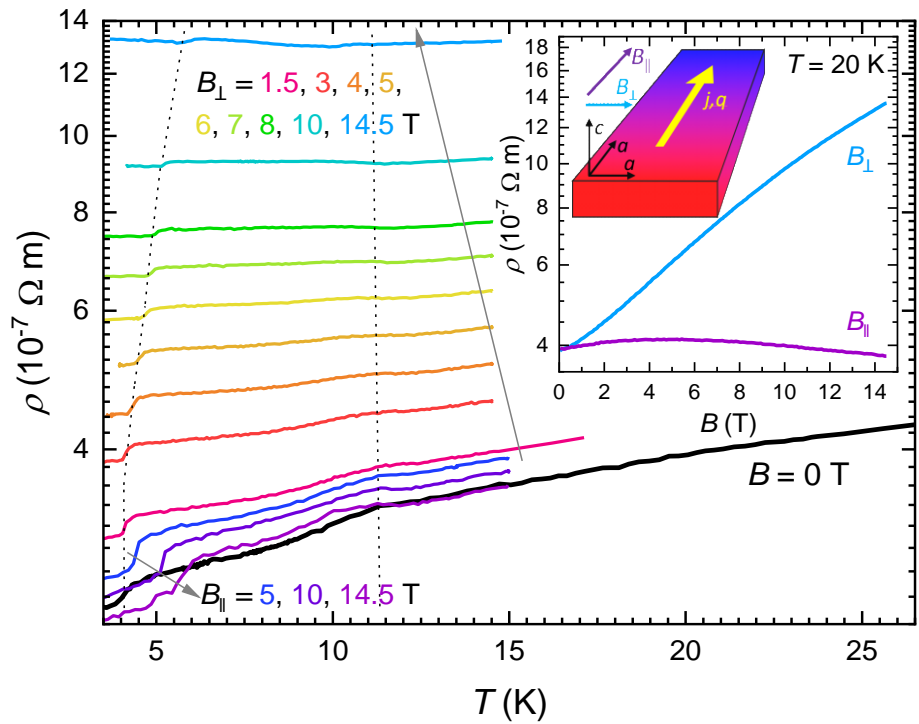

**Figure 1.** (Colour online) Temperature dependences of the electrical resistivity of SmAlSi measured at different magnetic field (from 0 to 14.5 T) in two configurations: perpendicular ($B_\perp \parallel a \perp j$) and parallel ($B_\parallel \parallel a \parallel j$). The dashed lines denote anomalies related to magnetic phase transitions. Inset presents schematic drawing of the vector configuration and the exemplary magnetic field dependences measured at $T = 20$ K. Both the main panel and inset are plotted on a semi-logarithmic scale.

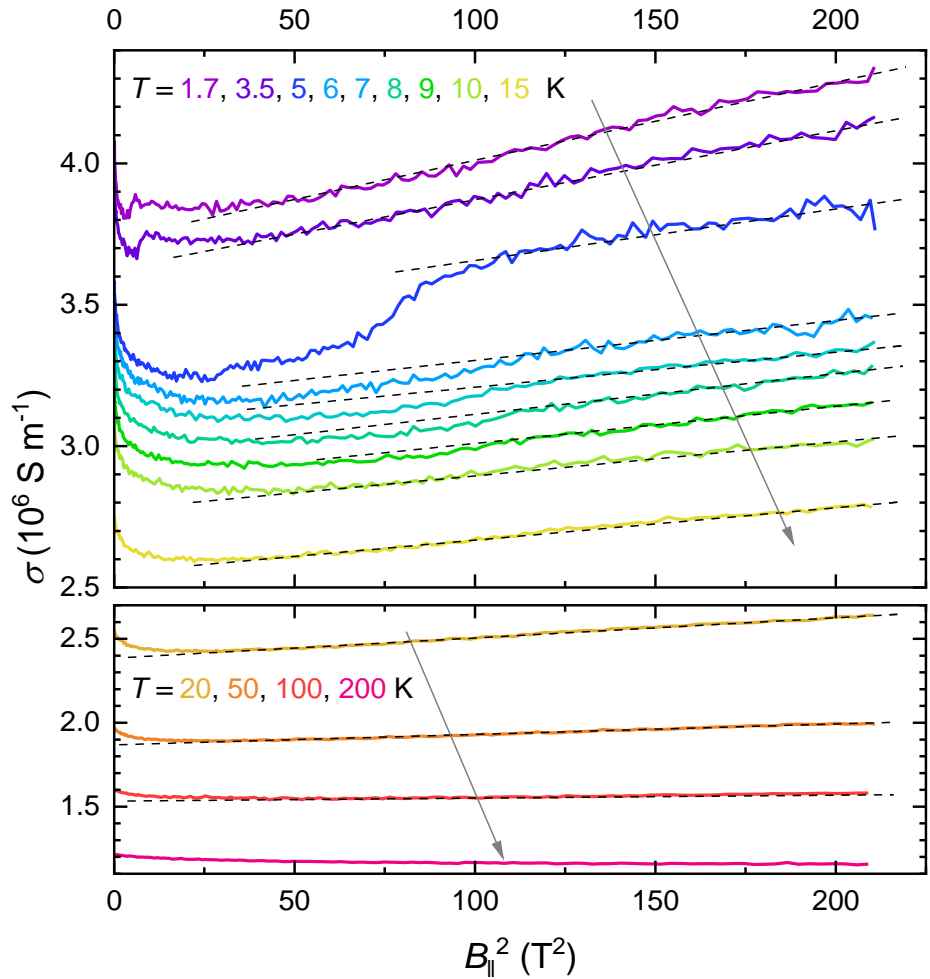

**Figure 2.** (Colour online) The electrical conductivity of SmAlSi for $B_\parallel \parallel a \parallel j$ plotted versus $B_\parallel^2$ for various temperatures. The dashed lines are to indicate $B_\parallel^2$ behaviour of $\sigma$ in the high field limit.

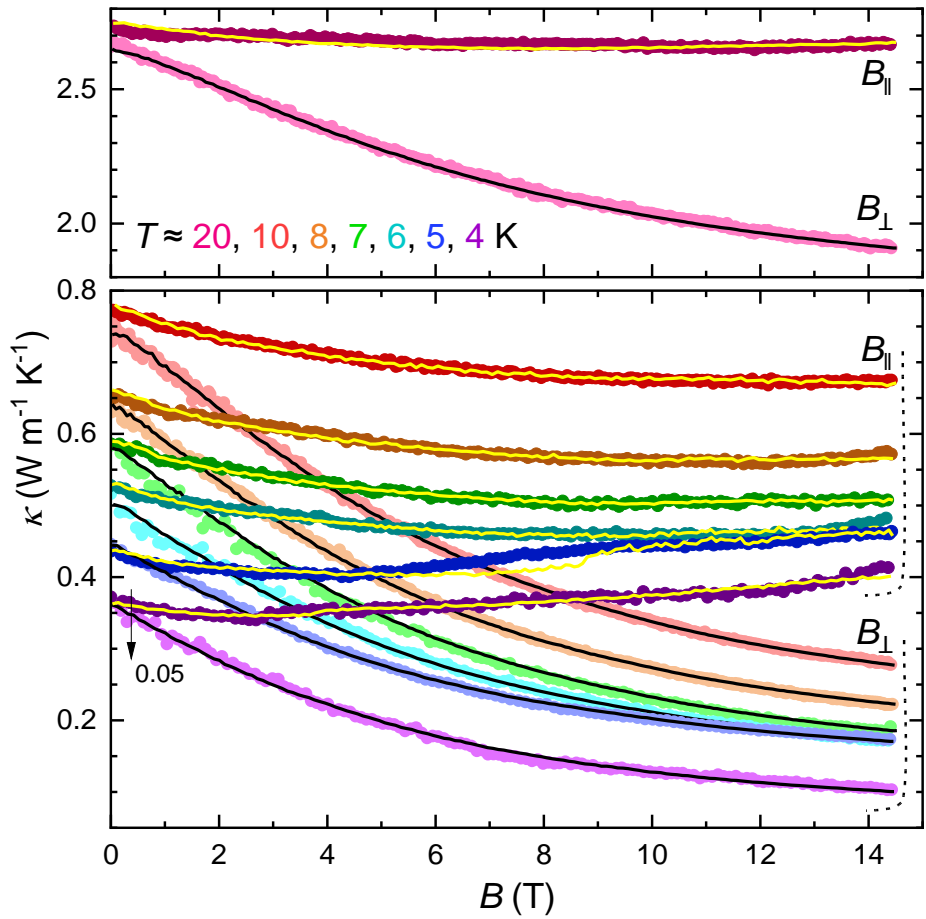

**Figure 3.** (Colour online) The magnetic field dependences of the thermal conductivity of SmAlSi. Light-coloured points present $\kappa(B_\perp)$ measured for $B_\perp \parallel a \perp \nabla T$ and dark coloured points $\kappa(B_\parallel)$ measured for $B_\parallel \parallel a \parallel \nabla T$. These dependences are superimposed with the thermal conductivity calculated from the electrical conductivity measurements for $B_\perp$ and $B_\parallel$ (black and yellow lines, respectively. The data for $T \approx 4$ K are shifted down by 0.05 W m$^{-1}$ K$^{-1}$ for the sake of clarity.

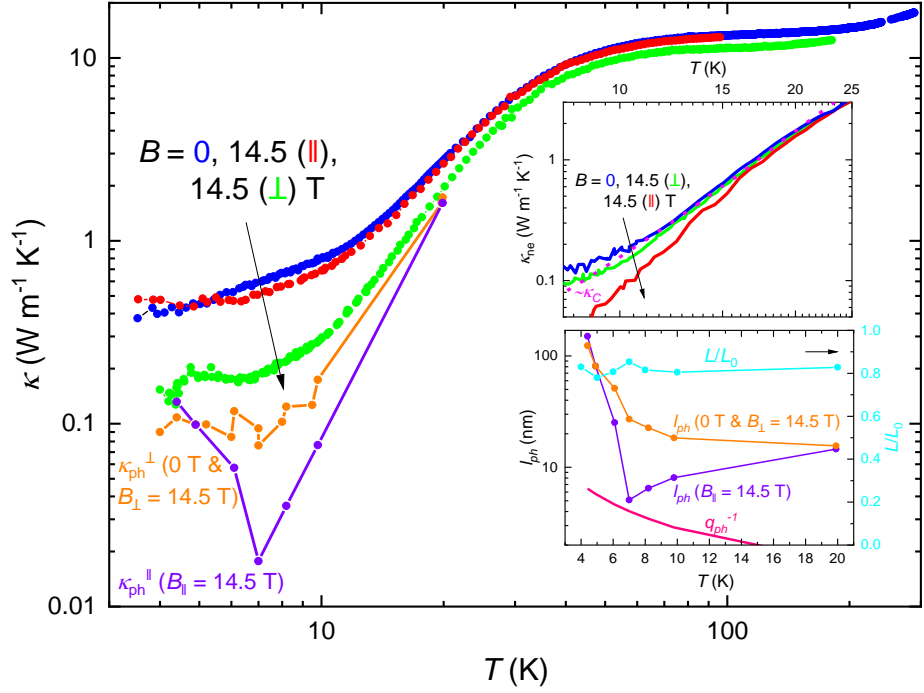

**Figure 4.** (Colour online) The temperature dependences of the thermal conductivity of SmAlSi for $B = 0$ T, $B_\perp = 14.5$ T, and $B_\parallel = 14.5$ T shown in the logarithmic scale. The main panel also presents temperature dependences of the phononic thermal conductivity for $B = 0$ T, $B_\perp = 14.5$ T, and $B_\parallel = 14.5$ T. The upper inset shows the non-electronic contribution to the thermal conductivity (solid lines) compared with the lattice thermal conductivity estimated from the specific heat (dashed line). The bottom inset presents the temperature dependences of the phonon mean free path for $B = 0$ T, $B_\perp = 14.5$ T, and $B_\parallel = 14.5$ T compared with the inverse of the wave vector of thermally excited phonons (left axis). There is also shown he normalised temperature dependence of the Lorenz number (right axis).

# Supplemental Material

**Anisotropic suppression of the phononic thermal conductivity by magnetic field in SmAlSi**

Mujeeb Ahmad [1], Md Shahin Alam [1], Xiaohan Yao [2], Fazel Tafti [2], Marcin Matusiak [2,3]

4. *International Research Centre MagTop, Institute of Physics, Polish Academy of Sciences, Aleja Lotnikow 32/46, PL-02668 Warsaw, Poland*
5. *Department of Physics, Boston College, Chestnut Hill, Massachusetts, 02467, USA*
6. *Institute of Low Temperature and Structure Research, Polish Academy of Sciences, ulica Okólna 2, 50-422 Wrocław, Poland*

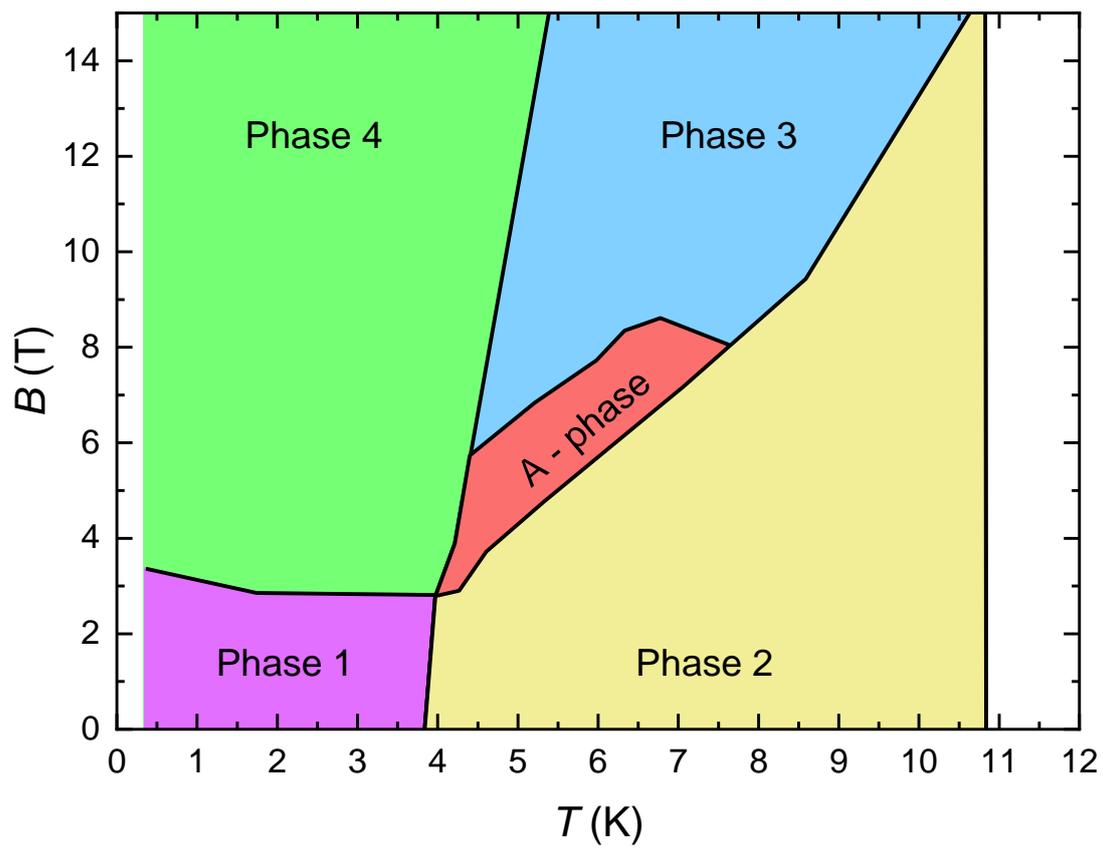

**Figure S1.** The schematic $B$ - $T$ phase diagram of SmAlSi based on the results from Ref. [1].